\documentclass{article}

\usepackage{PRIMEarxiv}

\usepackage[utf8]{inputenc} 
\usepackage[T1]{fontenc}    
\usepackage{hyperref}       
\usepackage{url}            
\usepackage{booktabs}       
\usepackage{amsfonts}       
\usepackage{nicefrac}       
\usepackage{microtype}      
\usepackage{lipsum}
\usepackage{fancyhdr}       
\usepackage{graphicx}       
\usepackage{caption}
\usepackage{subcaption}
\usepackage{xcolor}
\usepackage{amsmath}
\graphicspath{{media/}}     

\title{Comparison of Pebble Bed Velocity Profiles Between High-Fidelity and Intermediate-Fidelity Codes
\thanks{\textit{\underline{Citation}}: 
\textbf{Authors. Title. Pages.... DOI:000000/11111.}} 
}

\author{
  David Reger, Elia Merzari \\
  Pennsylvania State University \\
  University Park, PA 16803\\
  \texttt{dzr5281@psu.edu ebm5351@psu.edu} \\
   \And
  Paolo Balestra, Sebastian Schunert \\
  Idaho National Laboratory \\
  Idaho Falls, ID 83415\\
  \texttt{paolo.balestra@inl.gov sebastian.schunert@inl.gov} \\
  \And
  Yassin Hassan \\
  Texas A\&M University  \\
  College Station, TX 77843 \\
  \texttt{y-hassan@tamu.edu}
}

\begin{document}
\maketitle

\begin{abstract}
Recent  interest for the development of high-temperature gas reactors has increased the need for more advanced understanding of flow characteristics in randomly packed pebble beds. A proper understanding of these flow characteristics can provide a better idea of the cooling capabilities of the system in both normal operation and accident scenarios. In order to enhance the accuracy of computationally efficient, intermediate fidelity modeling, high-fidelity simulation may be used to generate correlative data. For this research, NekRS, a GPU-enabled spectral-element computational fluid dynamics code, was used in order to produce the high-fidelity flow data for beds of 1,568 and 45,000 pebbles. Idaho National Lab’s Pronghorn porous media code was used as the intermediate fidelity code. The results of the high-fidelity model were separated into multiple concentric regions in order to extract porosity and velocity averages in each region. The porosity values were input into the Pronghorn model and the resulting velocity profile was compared with that from NekRS. Both cases were run with a Reynolds number of 20,000 based on pebble diameter. The Pronghorn results were found to significantly overestimate the velocity in the outermost region indicating that changes in the porosity alone do not cause the difference in fluid velocity. We conclude that further work is necessary to develop a more effective drag coefficient correlation for the near-wall region and improve predictive capabilities of intermediate fidelity models.
\end{abstract}


\section*{INTRODUCTION}
A number of generation IV reactor systems are currently in development that make use of pebble fuel rather than traditional fuel rods. In these systems, the uranium fuel is embedded in graphite pebbles that are held in the reactor vessel. The coolant of these designs is typically either an inert gas, such as helium, or a liquid, such as a molten salt. In order to advance the deployment of these pebble bed systems, a better understanding of the flow in randomly packed beds is necessary, particularly in the near-wall region where the system may experience bypass flow as flow escapes through gaps between graphite blocks that make up the reactor vessel. This phenomena has been identified of being of high importance, but unfortunately is not well understood \cite{PIRT}.

The principle effect of interest for this research is known as the wall-channel effect. The wall-channel effect is a phenomenon observed in packed beds where the porosity increases sharply near the wall of the bed as the wall disrupts the bed geometry. As a result of this effect, the velocity of a fluid flowing through a packed bed, such as in a pebble bed reactor, varies largely with this porosity change. These velocity changes can have a significant effects on the distribution of coolant flow in the core as well as on the heat transfer properties at the wall. The diversion of flow into this region of high porosity can lead to other hot parts of the core not being adequately cooled, and thus it is an area of concern for developing reactors.

In order to predict this diversion of flow into the outer region, one must be able to predict the pressure drop throughout the system. Multiple correlations have been developed in the past in order to attempt this prediction of the pressure drop. Ergun \cite{Ergun} was one of the first contributors to suggest a pressure drop correlation for use in packed backs, but his correlation did not include any inclusion of wall-effects. Riechelt \cite{Reichelt} later conducted numerous experimental studies to improve Ergun's correlation in order to account for a larger range of Reynolds numbers, porosities, and bed-to-pebble diameter ratios. Felice and Gibilaro \cite{Felice} suggested a further improvement to the Ergun correlation by introducing a simple model that addressed the pressure drop only in the near-wall region. This model was then later improved by Cheng \cite{Cheng} who parameterized the model for a range of pebble to bed diameter ratios. Eisfeld and Shnitzelein \cite{Eisfeld} also proposed improvements to Riechelt's correlation by analyzing data available in literature. The German Nuclear Safety Standards commission (KTA) \cite{KTA} have also introduced correlations for applications to high-temperature gas reactors. This correlation will be focused on for this study, due to this specific application to HTGR's. More recently, experimental work has been performed by Hassan and Kang \cite{HassanKang} in order to develop a new pressure drop correlation for systems with a high Reynolds number and a low bed-to-pebble diameter ratio. Additionally, Nguyen et. al. \cite{Nguyen} have performed an experimental study in order to analyze cross-flow mixing

Computational studies have also been performed to analyze the flow in packed beds, along with the bypass-flow phenomenon. Atmakidis and Kenig \cite{kenig} have conducted a CFD study analyzing the wall effects in both regular and irregular packed beds in order to compare results to available correlations. Das et. al \cite{Das} have also performed a DNS study in order to simulate the flow and heat transfer through a randomly packed bed for a variety of Reynolds numbers and pebble-to-bed diameter ratios. Bypass flows have been simulated by Jun et. al. \cite{Jun} using CFD to calculate bypass flow rates in the HTR-PM high-temperature gas reactor. Yildiz et. al. \cite{Yildiz} have also performed DNS simulations of a pebble bed to compare DNS data to a number of the aforementioned correlations.

This work aims to contribute to the development of a pressure drop correlation for packed beds that can accurately represent the pressure drop in the near-wall region for HTGR's and FHR's. This may then be used to better predict bypass flow and identify potential core hot spots.

In particular we examine several potential modifications to the porous media formulation as implemented in Pronghorn ~\cite{NOVAK2021107968}, an intermediate fidelity code in development at Idaho National Laboratory. We use novel large-scale simulation results obtained with the spectral element Nek5000/NekRS as benchmark data . Detailed comparisons between the two codes are provided.

\section{METHODS}
This section presents the methods used for this study, including a background on the two codes used, along with the discretization of the computational domain
\subsection{NEKRS}
NekRS was chosen as the high-fidelity code, as it's GPU compatibility allows it to run the large pebble bed simulation with relative ease. 
NekRS is a new GPU-oriented version of Nek5000 \cite{fischer2016}, an established open-source  spectral element code in development at Argonne National Laboratory. NekRS is also capable of running on CPU's. It represents a significant redesign of the code. While written in C++, NekRS is able to link to Nek5000 to leverage its extensive pre- and post- processing utilities. NekRS has been built primarily under the auspices of the DOE’s Exascale Computing Project. It realizes high throughput on advanced GPU nodes and demonstrates excellent scalability. Details of NekRS performance for nuclear applications have been provided in a recent publication \cite{merzari2020}. 

Two cases were created and simulated with NekRS, one with 1,568 pebbles and one with 45,000 pebbles. The meshes for these cases were created using a novel Voronoi cell approach as part of the Cardinal multi-physics project \cite{Cardinal}.  This is a  tailored approach to the generation of all-hexahedral meshes in the pebble void regions that is based on a tessellation of the Voronoi cells defined by the pebble centers. The work is discussed in  detail at Merzari et al. \cite{merzari2021}.

Pebbled centers were generated performing a Discrete Element Method (DEM) calculation with a target porosity of 40\%. An example of a center-plane slice of the 1,568 pebbles bed can be seen in Fig. \ref{PronghornMesh}. For the simulation, the cases were run with an average Reynolds number of 20,000 based on the pebble diameter. Both cases were run on Oak Ridge National Lab's Summit Supercomputer in order to utilize the parallel-GPU capabilities of the code. The 1,568 pebble case was run across 72 GPU's with a timestep of $5.0*10^{-4}$s and the 45,000 pebble case was run across 1,788 GPU's with a timestep of $3.0*10^{-4}$s. Both cases were run for roughly 100,000 time steps, with time averaging occurring after the first 20,000 time steps. The turbulence modeling option employed is Large Eddy Simulation, with an explicit filter mimicking the effect of dissipation of the sub-grid scales. Once results were averaged, a number of postprocessing actions, discussed in section 1.4, were taken.

\subsection{PRONGHORN}
The intermediate-fidelity code of choice for this study was Idaho National Lab's Pronghorn that implements several porous media formulations~\cite{NOVAK2021107968}. Pronghorn is a finite element thermal-hydraulics code that is built on the Multiphysics Object-Oriented Simulation Environment (MOOSE) framework. Pronghorn is intended to provide transient and steady state simulation results in short code execution times to assist in design scoping studies or to provide boundary conditions for system-level analysis of pebble bed systems.

Pronghorn is capable of multiple solver models in order to resolve the fluid and temperature distributions of a system, but the one chosen for this work is the porous media, compressible Navier-Stokes model. The strong form of this model is given by \cite{PronghornTheoryManual}:

\begin{equation} \label{eqn1}
  \epsilon\frac{\partial{\rho}}{\partial{t}}+\nabla\cdot(\epsilon\rho_{f}\vec{V})=0,
\end{equation}
\begin{equation}
 \begin{aligned}
  \epsilon\frac{\partial{(\rho_{f}\vec{V})}}{\partial{t}}+\nabla\cdot(\epsilon\rho_{f}\vec{V}\vec{V})+\epsilon\nabla{P} -\epsilon\rho_{f}\vec{g}+W\rho_{f}\vec{V} \\
  -\nabla\cdot(\mu\nabla\vec{V})=0,
 \end{aligned}
\end{equation}
\begin{equation}
 \begin{aligned}
  \epsilon\frac{\partial{(\rho_{f}{E}_{f})}}{\partial{t}}+\nabla\cdot(\epsilon{H}_{f}\rho_{f}\vec{V})-\nabla\cdot(\kappa_{f}\nabla T_{f})-\epsilon\rho_{f}\vec{g}\cdot\vec{V} \\
  +\alpha(T_{f}-T{s})-\dot{q_{f}} = 0,
 \end{aligned}
\end{equation}
\begin{equation}
  (1-\epsilon)\rho_{s}{C}_{p,s}\frac{\partial{T_{s}}}{\partial{t}}-\nabla\cdot(\kappa_{s}\nabla{T}_{s})+\alpha(T_{s}-T{f})-\dot{q_{s}} = 0.
\end{equation}

\begin{figure}
  \center
  \includegraphics[width=0.422\textwidth]{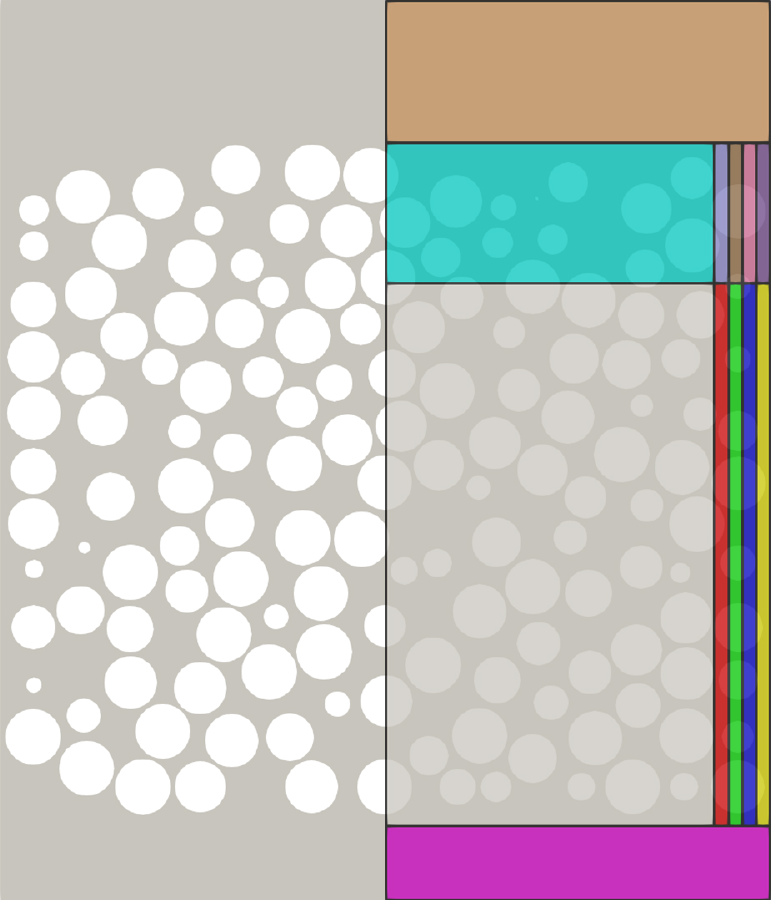}
  \caption{CENTERPLANE SLICE OF THE NEKRS CASE WITH THE R-Z PRONGHORN 5-RING CASE OVERLAYED ON THE RIGHT SIDE}
  \label{PronghornMesh}
\end{figure}

Two cases consistent with to the NekRS cases were created in Pronghorn to provide comparative data. In these cases, R-Z meshes were created that were nondimensionally identical to the NekRS cases.
Helium was selected as the coolant fluid, and the inlet velocity was scaled to achieve an average Reynolds number of 20,000. The pebble bed was separated into several regions, with each region assigned the porosity values that were calculated in the postprocessing step of the NekRS simulation. Additionally, the entrance and exit regions of the systems were included in order to gauge how well Pronghorn would be able to represent these regions and to ensure further similarity to the NekRS case. The mesh for the Pronghorn case can be found overlayed on top of the corresponding NekRS case for the 1,568 pebble system on the right of Fig. \ref{PronghornMesh}.

\begin{figure}
    \centering
    \includegraphics[width=0.47\textwidth]{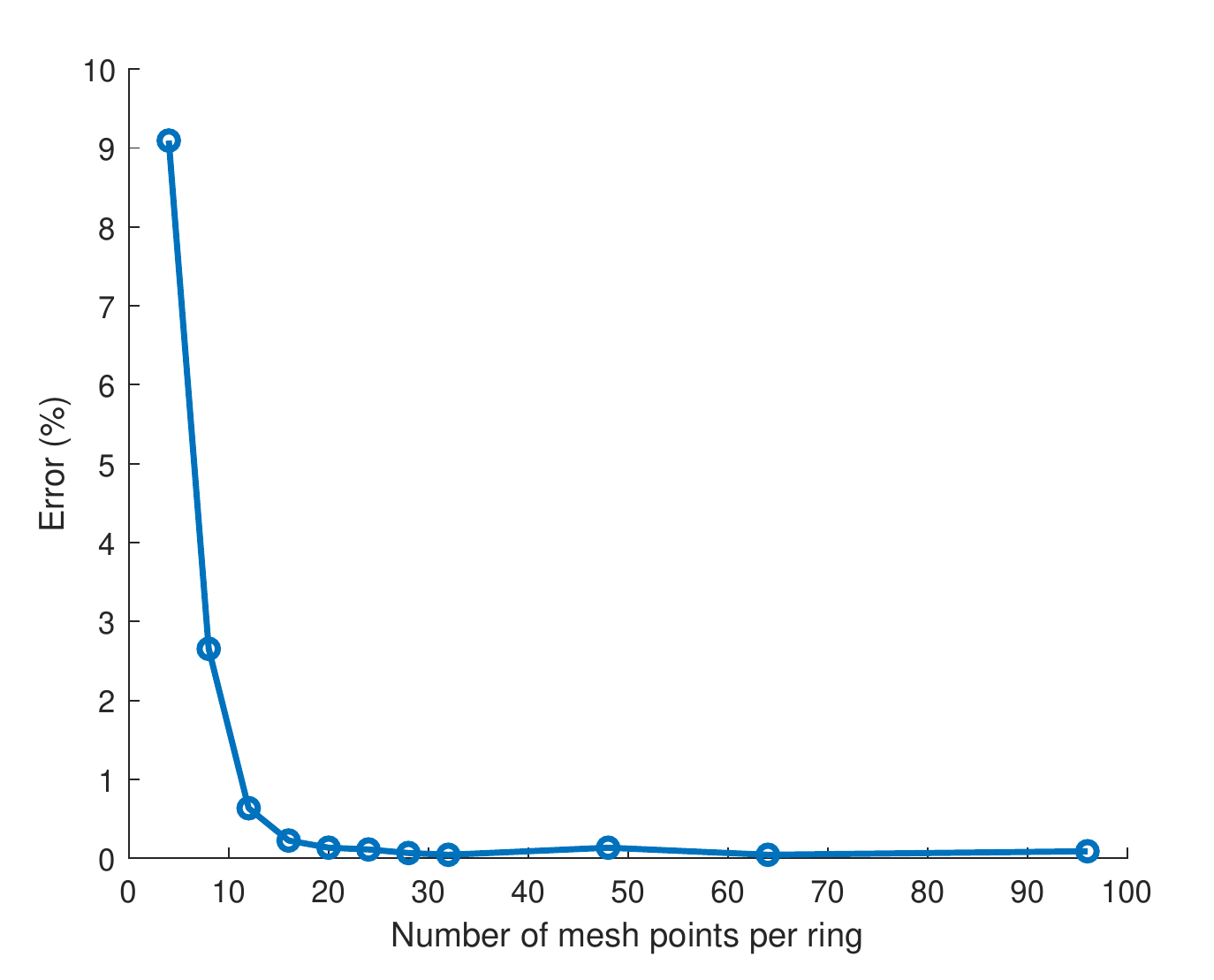}
    \caption{Error in the Pronghorn calculated average value in the outer ring of the 1,568 pebbles system versus number of radial mesh points used for each ring}
    \label{MCR}
\end{figure}

A mesh convergence investigation was performed on the Pronghorn cases to ensure that the mesh was sufficiently fine to accurately simulate the system. A plot of the error of the calculated average in the outer ring versus the number of radial mesh points used per ring can be found in Fig. \ref{MCR}. It can be seen that there is little improvement for more than 32 points, so 32 points were used to produce the results seen later.

\subsection{DRAG COEFFICIENT INVESTIGATION METHODOLOGY}
The goal of this study is to investigate Pronghorn's ability to capture the wall-channel effect if the porosity changes are manually implemented into each case. If it is unable to accurately reproduce the effect of wall-channeling on the velocity distribution, then a better pressure drop correlation must be developed. Once Pronghorn's capabilities are better understood, the drag coefficients in the high-porosity region near the wall may then be manually altered in order to produce results that match those seen from the LES data. Alternatively, it may be possible to compute the drag coefficients from the NekRS simulation data so that they can be input into Pronghorn without requiring additional user input. If these manual changes are to prove to be effective in matching the LES data more closely, additional simulation cases may be created to gather data for different bed sizes, pebble sizes, and Reynolds numbers. This data can then be used to train a reduced-order model and produce a correlation for the drag coefficients in the near-wall region. With this correlation, a Pronghorn object could be created in order to address wall-channeling for a single bed domain with specified bulk porosity, allowing the effect to be captured without increasing the complexity of the case for the user.

\subsection{SIMULATION APPROACH}
Investigation of the wall-channel effect in Pronghorn requires the pebble bed to be separated into multiple subdomains in order to represent the changes in porosity that occur near the wall. For this study, the domain was separated into multiple concentric subdomains, with one bulk subdomain in the center of the system, and several smaller subdomains located near the outside of the system in order to capture the wall-channel effect. This separation was overlayed on the NekRS cases in the postprocessing step, allowing for the extraction of the average porosity and velocity values for each of the rings. With the average porosity of each ring known, these values were then used as inputs in the equivalent Pronghorn cases to represent these various subdomains. An example of this representation can be seen in Fig. \ref{PronghornMesh}. It was originally unknown as to how many subdomains were necessary to capture the wall-channel effect, so multiple cases were created where the outer region of the system of width $D_{pebble}$ was separated into 4,2, and 1 subregions in order to gauge the sensitivity of this choice. Initially, the Pronghorn cases were run using the default implementation of the KTA drag correlation that is available in Pronghorn. The results from these runs were then compared with the NekRS results to determine the areas of improvement. Once these discrepancies were discovered, the drag coefficients in the outermost ring of the Pronghorn case were adjusted to reach better agreement with the NekRS results.
\section{ANALYSIS AND RESULTS}
This section provides the resulting velocity fields and profiles for the NekRS and Pronghorn cases. Additionally, it demonstrates the ability to improve the drag coefficients used in Pronghorn.
\subsection{NEKRS RESULTS}
\begin{figure}
  \center
  \includegraphics[width=0.7\textwidth]{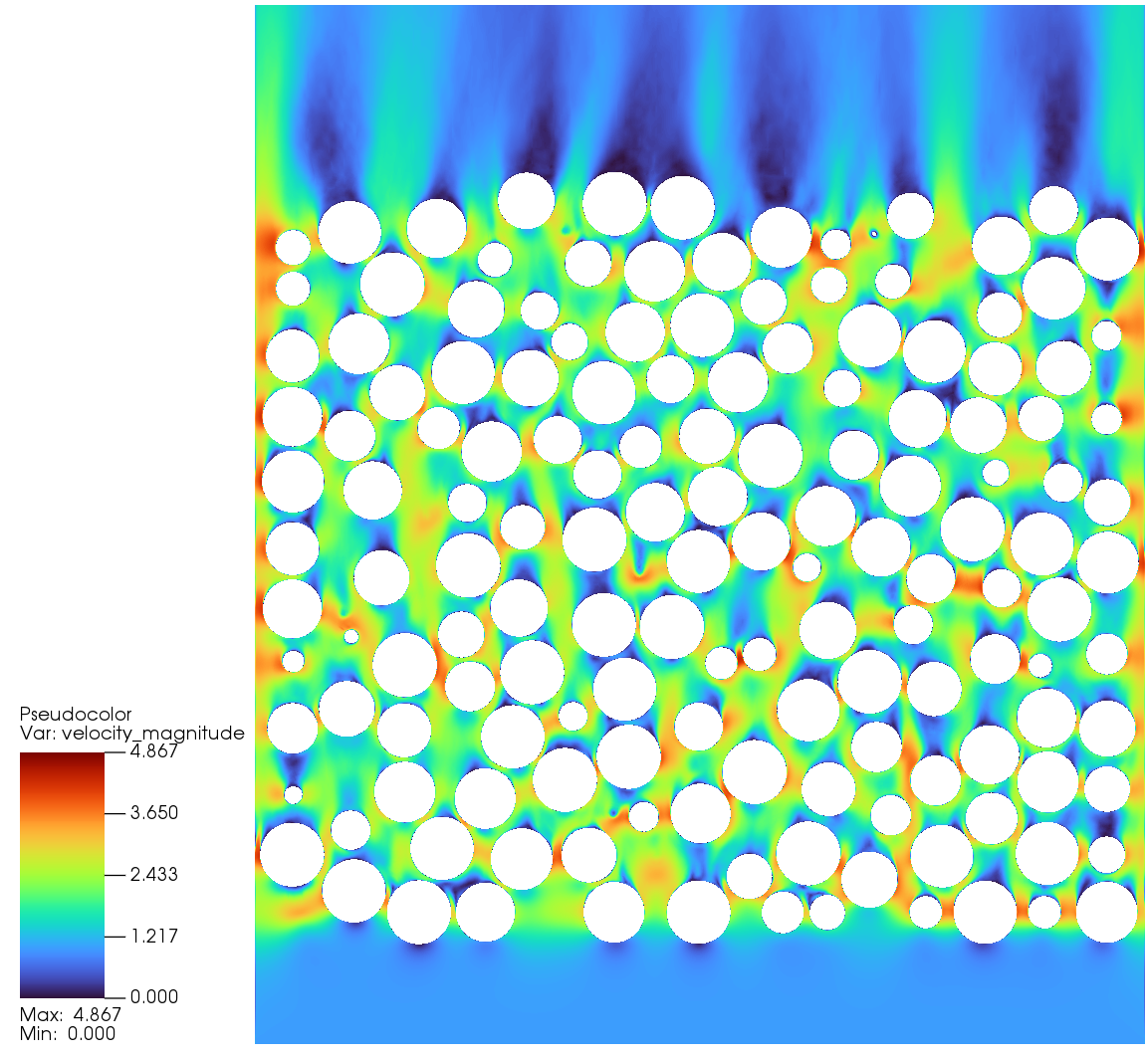}
  \caption{CENTERPLANE SLICE OF THE NEKRS VELOCITY FIELD FOR THE 1,568 PEBBLE CASE. NOTE ALL PEBBLES ARE THE SAME SIZE, BUT APPEAR DIFFERENT DUE TO THE SLICE FILTER USED.}
  \label{1568NekData}
\end{figure}

\begin{figure}
  \center
  \includegraphics[width=1\textwidth]{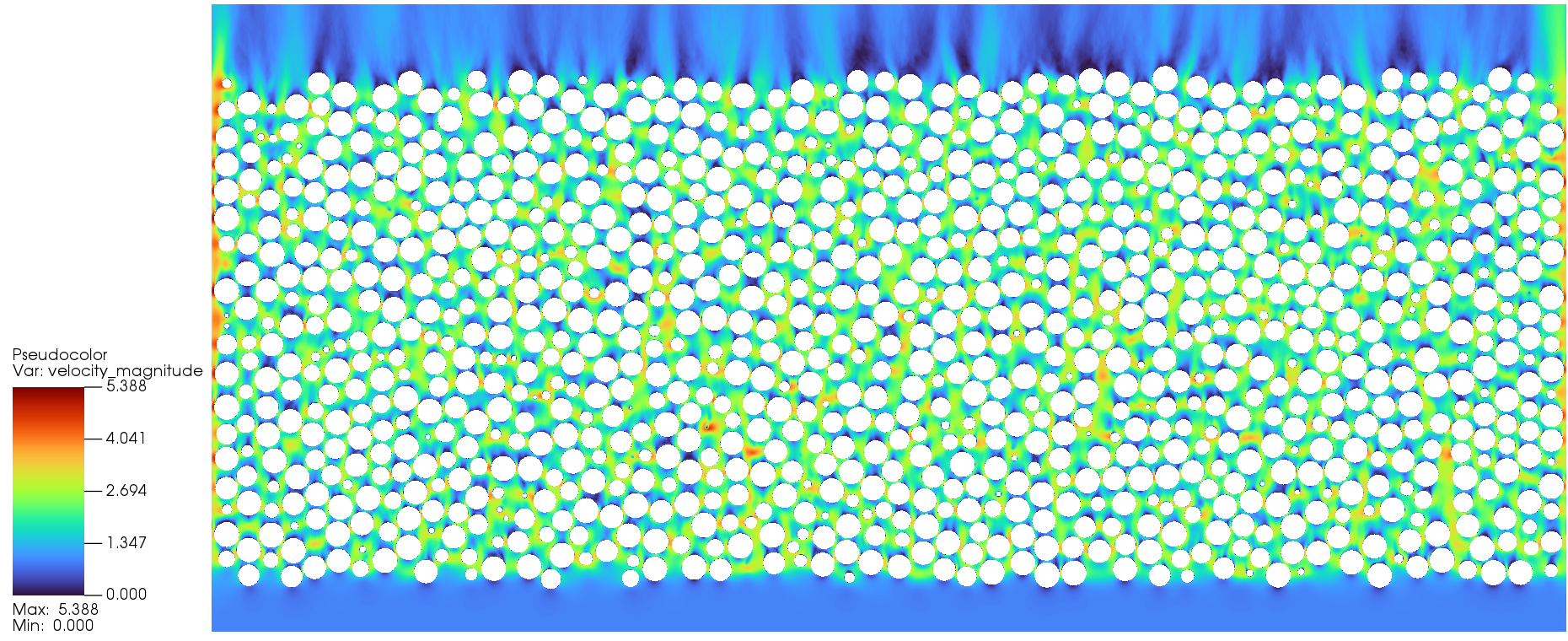}
  \caption{CENTERPLANE SLICE OF THE NEKRS VELOCITY FIELD FOR THE 45,000 PEBBLE CASE. NOTE ALL PEBBLES ARE THE SAME SIZE, BUT APPEAR DIFFERENT DUE TO THE SLICE FILTER USED.}
  \label{45kNekData}
\end{figure}

\begin{figure}
    \centering
    \includegraphics[width=0.7\textwidth]{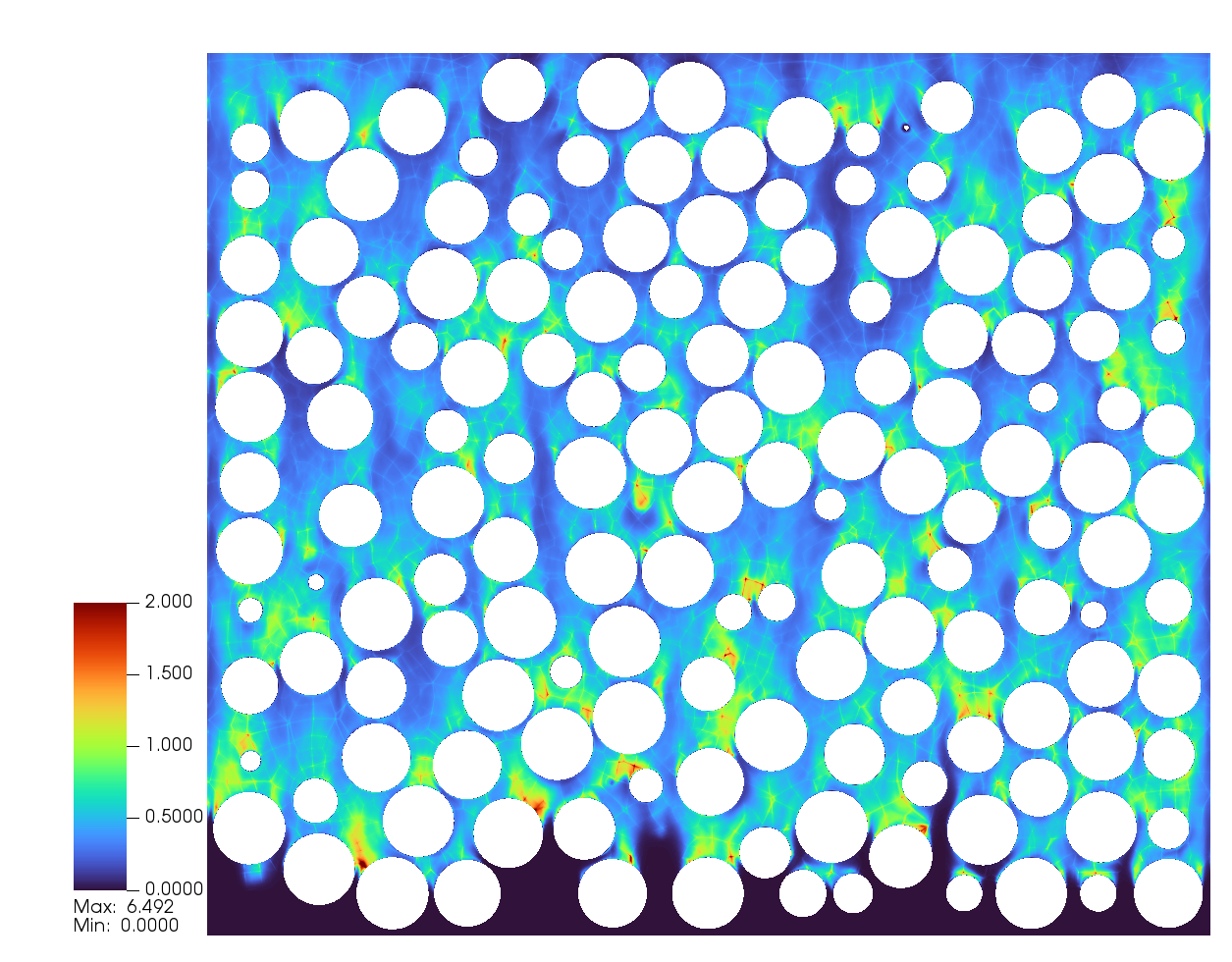}
    \caption{CENTERPLANE SLICE OF THE NEKRS TURBULENT KINETIC ENERGY FOR THE 1,568 PEBBLES CASE. ENTRANCE AND EXIT REGIONS ARE EXCLUDED.}
    \label{TKE}
\end{figure}

\begin{figure}
  \center
  \includegraphics[width=0.47\textwidth]{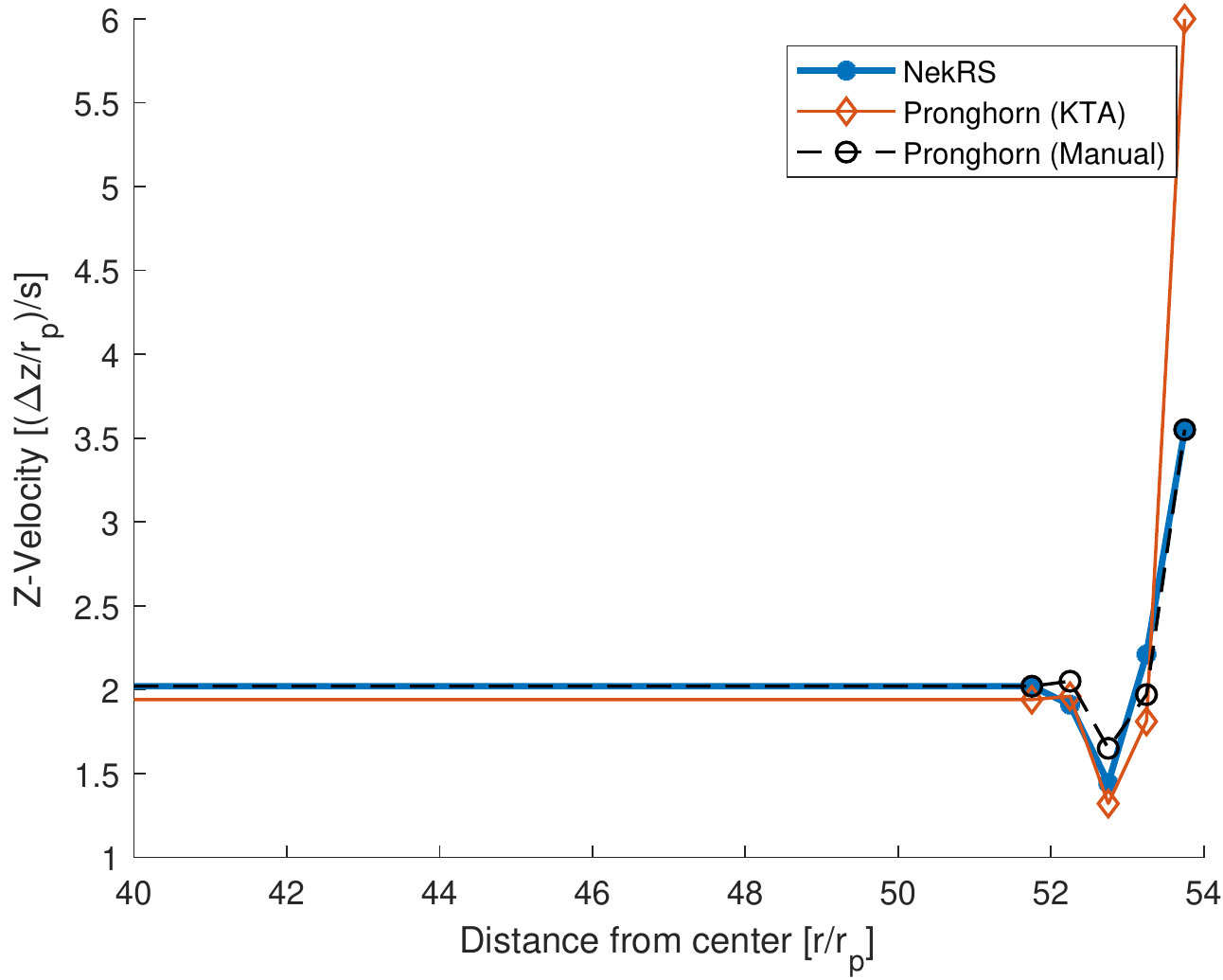}
  \caption{RADIAL PROFILE FOR THE STREAMWISE VELOCITY OF THE 45,000 PEBBLE CASE FOR NEKRS, PRONGHORN KTA, AND ADJUSTED PRONGHORN WITH 5 RING SUBDIVISIONS.}
  \label{45kRadial}
\end{figure}

The time-averaged velocity fields for the 1,568 and 45,000 pebble cases can be found in Fig. \ref{1568NekData} and \ref{45kNekData}. From these figures, it is possible to observe the increase in velocity that is seen in the near-wall region. It is also possible to observe the more orderly structure of the pebbles that are in contact with the wall. Postprocessing was performed on these time-averaged results in order to determine radial and axial velocity profiles for each of the regions. These results can be found in Figs. \ref{RadialProfiles} and \ref{45kRadial}. A visualization of the turbulent kinetic energy may also be found in Fig. \ref{TKE}. It should be noted that there is visible artifacting around the edges of elements in this visualization, indicating that the resolution of the mesh may need to be increased for further runs.


\begin{figure}
  \center
  \begin{subfigure}{0.47\textwidth}
    \center
    \includegraphics[width=\textwidth]{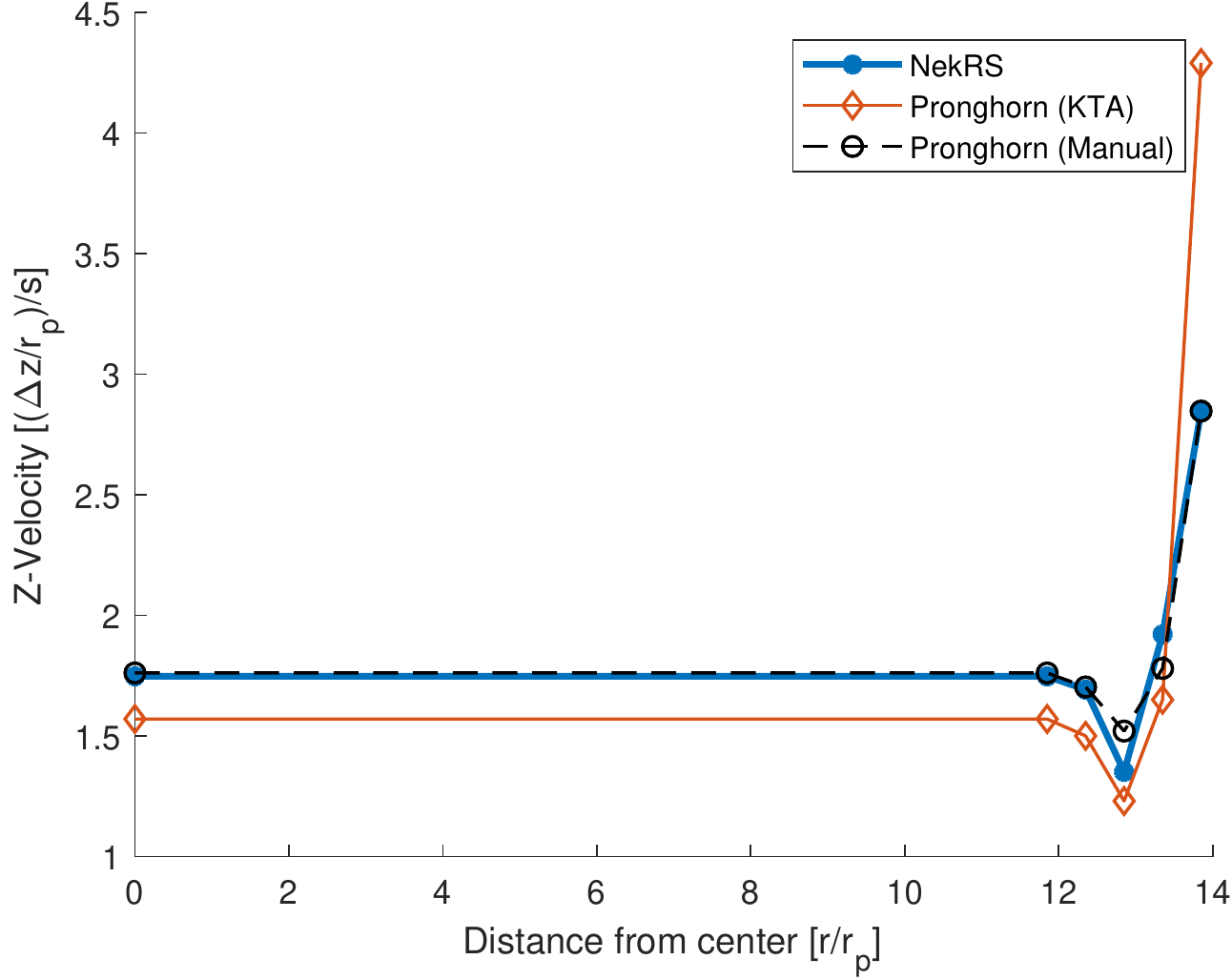}
    \caption{}
    \label{5Ring1568}
  \end{subfigure}
  \begin{subfigure}{0.47\textwidth}
    \center
    \includegraphics[width=\textwidth]{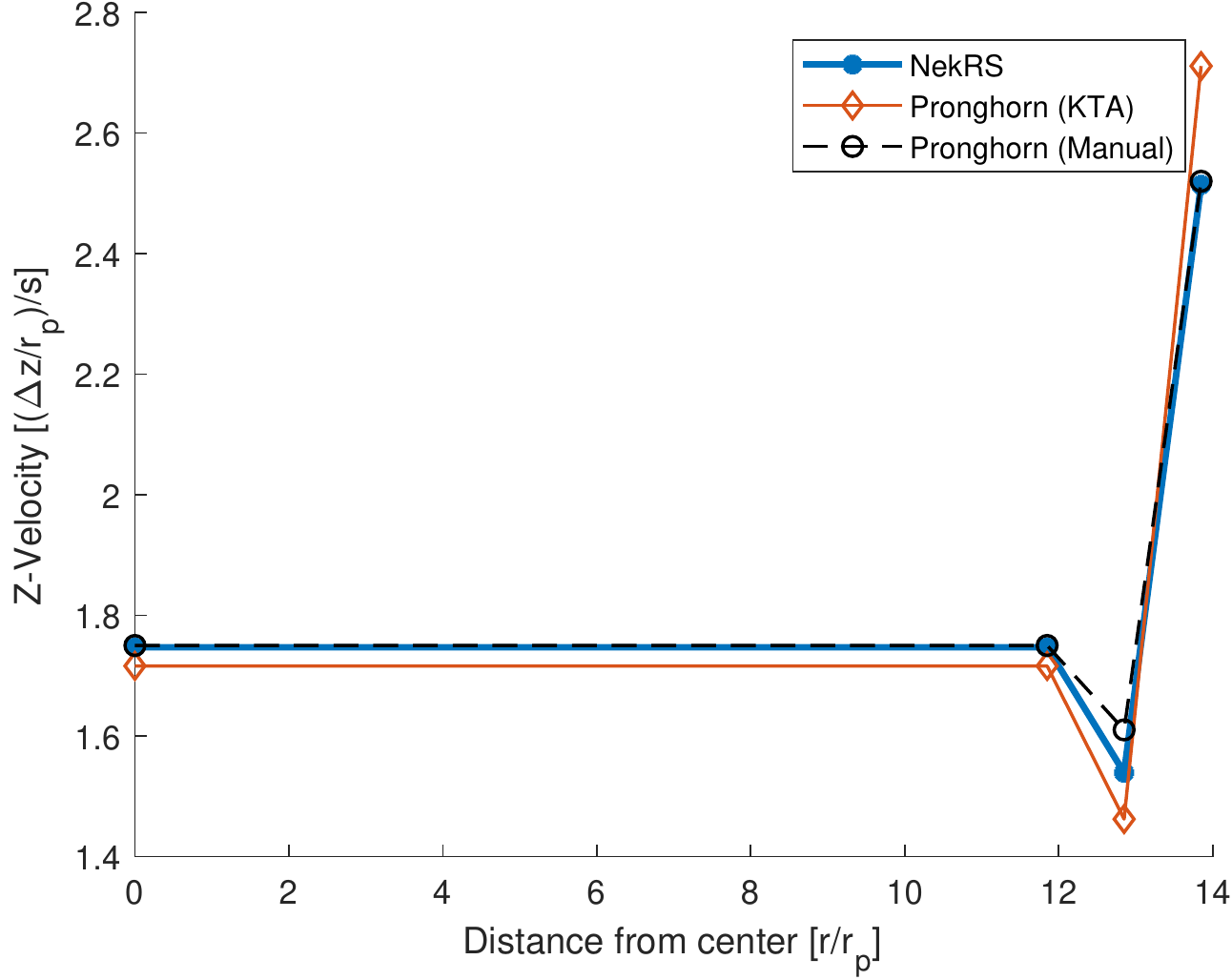}
    \caption{}
    \label{3Ring1568}
  \end{subfigure}
  \label{RadialProfiles}
\end{figure}

\begin{figure}
\continuedfloat
\center
  \begin{subfigure}{0.47\textwidth}
    \center
    \includegraphics[width=\textwidth]{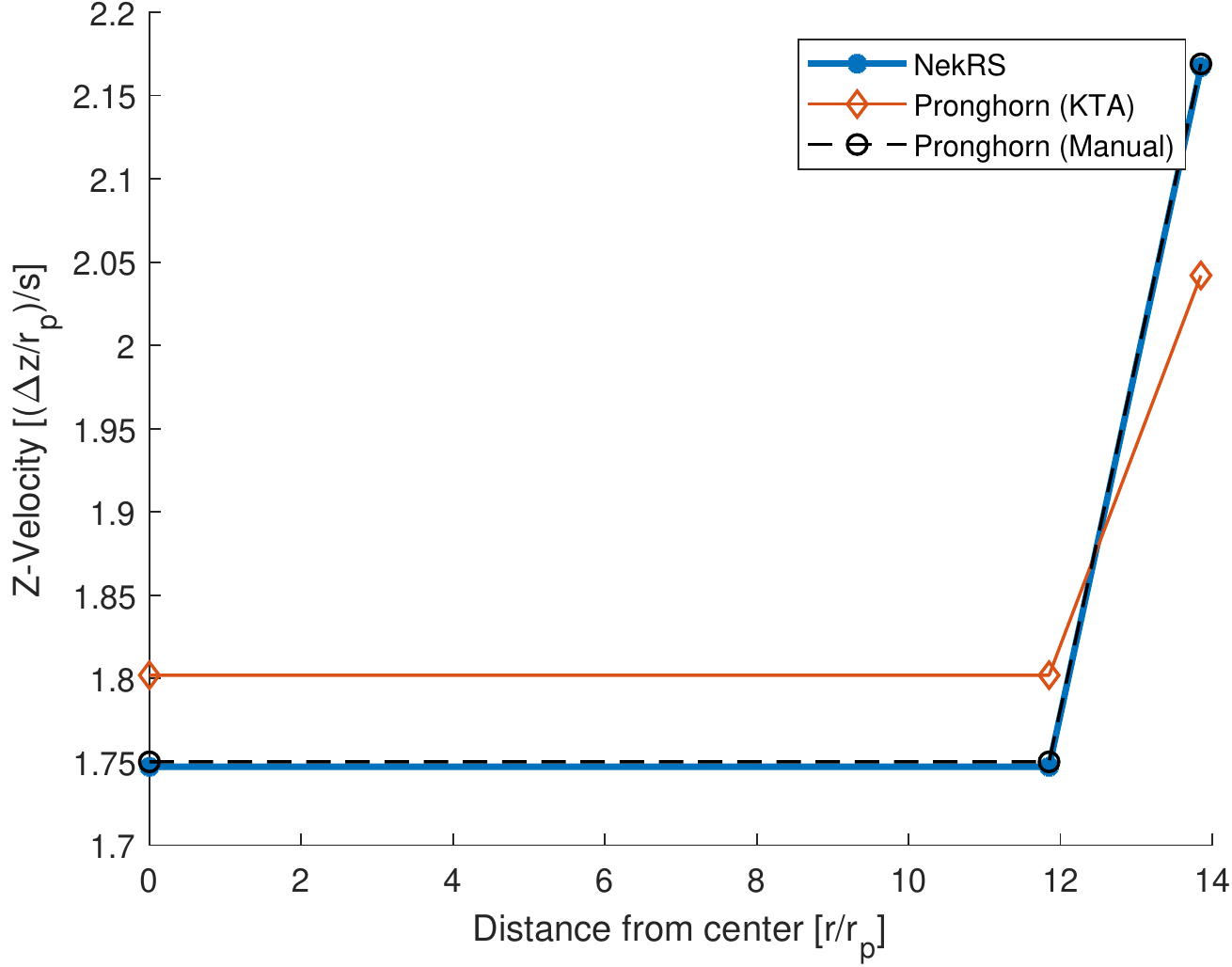}
    \caption{}
    \label{2Ring1568}
  \end{subfigure}
  \begin{subfigure}{0.47\textwidth}
    \center
    \includegraphics[width=\textwidth]{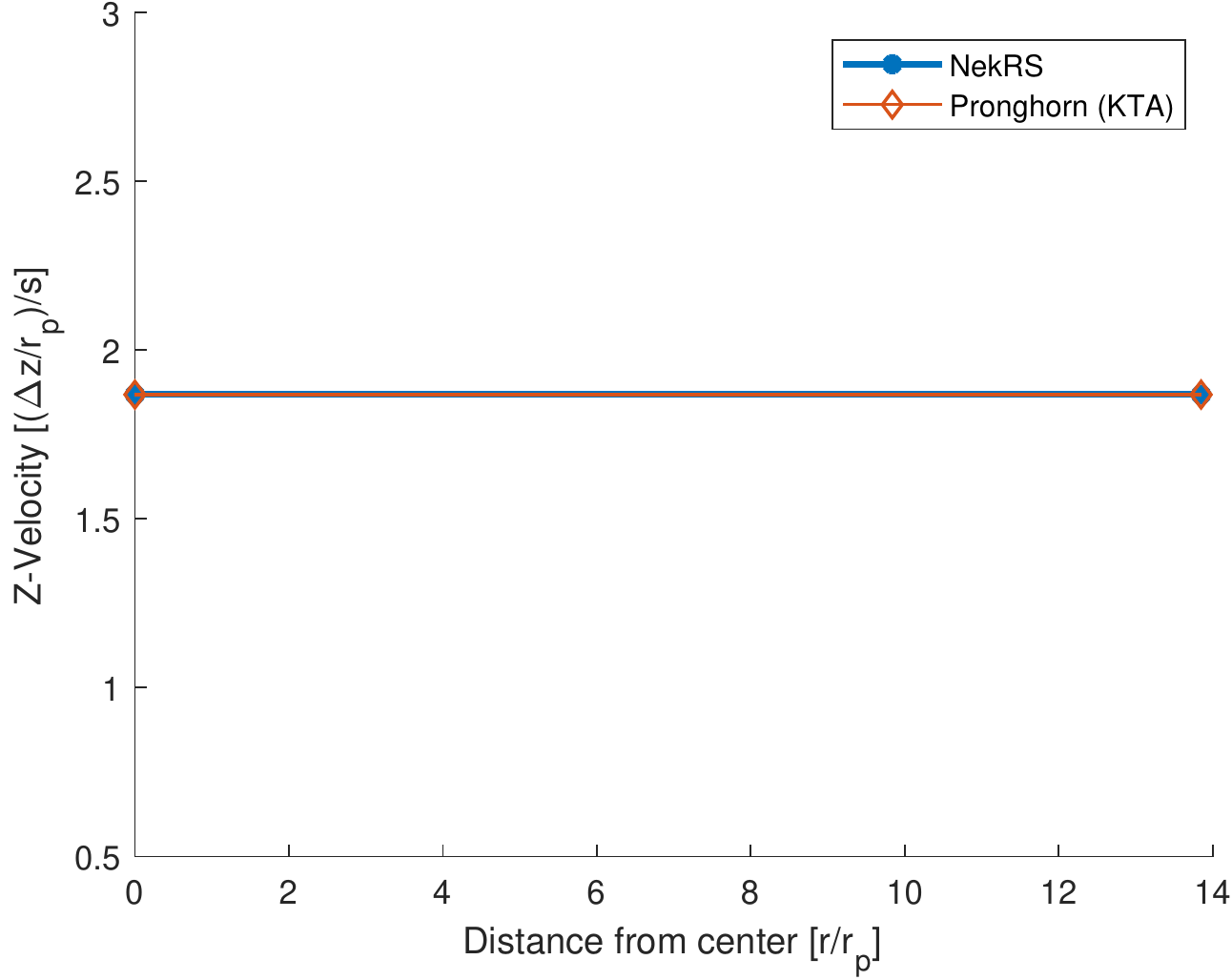}
    \caption{}
    \label{NoRing1568}
  \end{subfigure}
 \caption{RADIAL PROFILES FOR THE STREAMWISE VELOCITY OF THE 1,568 PEBBLES CASE FOR NEKRS, PRONGHORN KTA, AND ADJUSTED PRONGHORN FOR 5(a), 3(b), 2(c), AND 1(d) RING SUBDIVISIONS. }
 \label{RadialProfiles}
\end{figure}

\subsection{RADIAL VELOCITY PROFILE}
The radial profile of the streamwise velocity for the various cases can be seen in Fig. \ref{5Ring1568}-\ref{NoRing1568} for the 1,568 pebble system and Fig. \ref{45kRadial} for the 45,000 pebble system. From these figures, it can be seen that the KTA correlation does not accurately determine the drag coefficients in the outermost ring where the porosity changes significantly compared to the center of the system. The KTA correlation underestimates the drag in this region of high porosity, causing more of the flow to distribute to this ring. This is most prominently seen in the runs that separated the domain into 5 regions, as these runs show the high porosity of flow near the wall more clearly, meaning that the outer ring is much farther outside the range of valid porosities for the KTA correlation compared to the other cases. As a result, the cases of 5-rings will be pursued for further analysis, as they most prominently capture the wall-channel effect with no significant increase in complexity or code execution time. Additionally, the discrepancies appear to be relatively consistent between the 1,568 and 45,000 pebble cases, indicating that changing the porosity alone is not enough to reproduce the wall-channel effecting and the correlation must be improved. The case of a single domain of average porosity was also run, and it can be seen from Fig. \ref{NoRing1568} that the velocity matches the NekRS case entirely with the KTA correlation. This is expected, as reducing to a single region simplifies the problem so that correct prediction of the velocity requires only to satisfy the mass balance; this simplified setup served to show that the two cases were indeed consistent with each other. 

In order to demonstrate the feasibility of adjusting the drag coefficient in the outermost region, the drag coefficients in the outermost ring were manually adjusted while KTA values were used for all other rings. The coefficients were adjusted until the average velocity of the outer ring matched the result from NekRS. It can be seen that this change greatly improved the radial velocity distribution as a whole, as the lower distribution of flow into the outer ring allowed the other rings to receive more flow and match the NekRS results much more closely. There were still discrepancies between Pronghorn and NekRS in some of the inner rings that could be subject to further improvements.

\begin{figure}
  \includegraphics[width=0.47\textwidth]{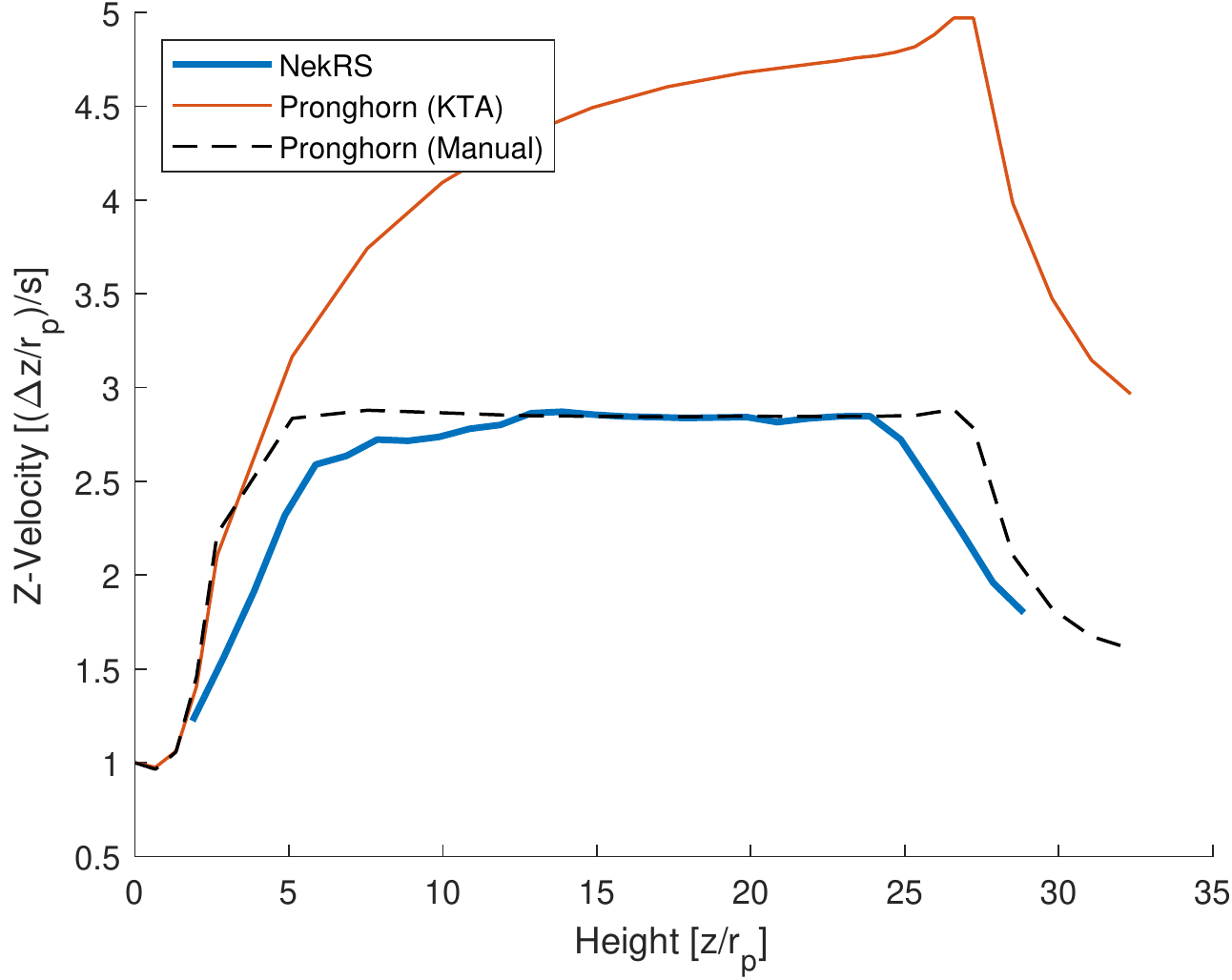}
  \centering
  \caption{AXIAL Z-VELOCITY PROFILE IN THE OUTERMOST RING FOR THE NEKRS, KTA PRONGHORN, AND ADJUSTED PRONGHORN CASES WITH THE 1,568 PEBBLE CASE}
  \label{1568Axial}
\end{figure}
\begin{figure}
  \center
  \includegraphics[width=0.47\textwidth]{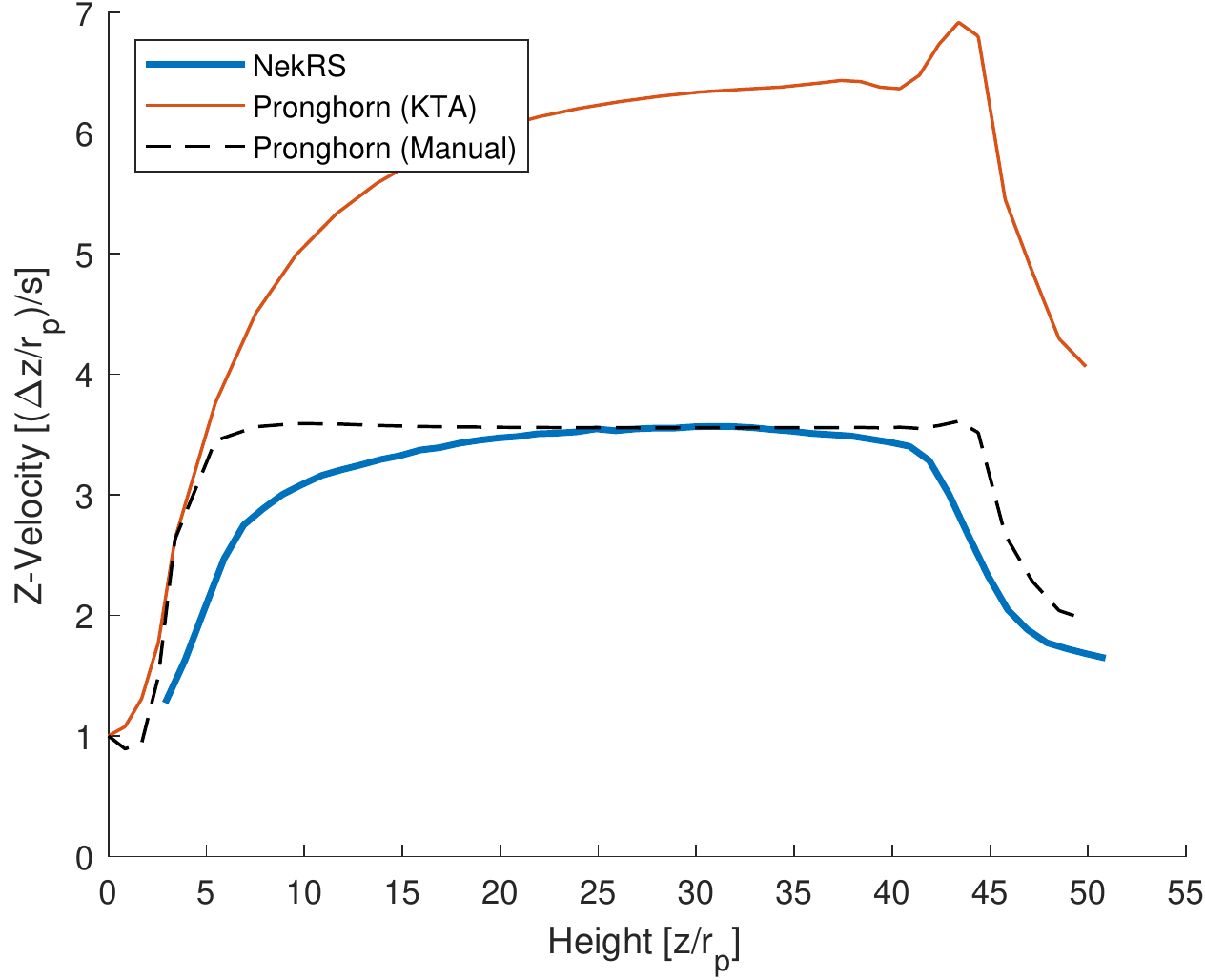}
  \caption{AXIAL Z-VELOCITY PROFILE IN THE OUTERMOST RING FOR THE NEKRS, KTA PRONGHORN, AND ADJUSTED PRONGHORN CASES WITH THE 45,000 PEBBLE CASE}
  \label{45kAxial}
\end{figure}
\subsection{AXIAL VELOCITY PROFILE}
The axial velocity profile for the outermost ring is shown in Fig. \ref{1568Axial} and \ref{45kAxial}. It can be seen that the Pronghorn case with the initial KTA correlation required a significantly longer distance to redistribute the flow from the entrance region to the various rings. The difference between the redistribution distances in the NekRS and Pronghorn cases was initially higher than an order of magnitude, indicating a significant deficiency in the standard porous media modeling approach for pebble bed reactors. The cause for this discrepancy was that the drag coefficient computed with the KTA correlation was applied isotropically and not just in the streamwise direction. A potential solution to this issue was to declare separate drag coefficients for each axis and setting the spanwise coefficients to zero. It can be seen that with this improvement, the Pronghorn case is capable of much more closely matching the redistribution distance observed in the NekRS case. This condition, however, is nonphysical, and finer tuning may be done in order to improve this agreement further.

\section{CONCLUSION}
In this study, the flow in a randomly packed pebble bed was analyzed using NekRS, a high-fidelity LES code, and Pronghorn, an intermediate-fidelity porous media code. Two cases were created with NekRS in order to simulate the flow through randomly packed beds of 1,568 and 45,000 pebbles at an average Reynolds number of 20,000. Postprocessing was performed on these two cases to separate the domains into multiple concentric subdomains, with the majority of the subdomains located near the outside wall of the system to observe the wall-channeling effect. For the 1,568 pebble case, cases were created with 5, 3, and 2 subdomains, along with a case with no subdomain separation. Average porosity values for each subdomain were calculated from the NekRS results and were used as inputs to the equivalent Pronghorn cases. Radial and axial velocity distributions were analyzed, and it was determined that using 5 subdomains was most effective at capturing the wall-channel effect. We note that these multi-region models do not incur any significant increase in case complexity or run time. 
In all  multi-region models, we observe also a general ineffectiveness of the KTA drag correlation in the wall-channel region. In fact, with the default KTA implementation in Pronghorn, the velocity in the outermost ring was significantly overestimated regardless of the number of regions employed, and the flow took much longer to distribute amongst the rings near the entrance region. 
We concluded that simply changing porosity is not sufficient to achieve accurate results. A different implementation of the drag coefficients was then used, whereby KTA values were used in the streamwise direction while spanwise drag coefficients were set to zero. This change improved the redistribution distance of the Pronghorn case to more closely match the NekRS result, although further investigation may allow for this to be improved by altering the spanwise drag coefficients. In order to attempt to demonstrate the feasibility of the method of addressing wall-channelling in porous media codes, the streamwise drag coefficients in the outermost ring of the Pronghorn cases were manually adjusted such that the average velocity of the outermost ring would be identical to the NekRS case. KTA values were still used for all other rings. Although there were still some discrepancies after these changes, there was significant improvement to the radial and axial velocity distributions, indicating that the method may be sufficient for representing the wall-channel effect in Pronghorn. Additional work will be performed in order to run cases with different sizes and Reynolds numbers in gather sufficient data to create a correlation for representing the wall-channel effect in porous media codes. Future work will also be dedicated to the investigation of heat transfer in the wall-channel region.

\bibliographystyle{unsrt}  
\bibliography{references}

\begin{thebibliography}{10}

\bibitem{PIRT}
S.J. Ball.
\newblock {Next Generation Nuclear Plant Phenomena Identification and Ranking
  Tables (PIRTs)}.
\newblock Technical report, NRC, 2007.

\bibitem{Ergun}
S.~Ergun.
\newblock Fluid flow through packed columns.
\newblock {\em Journal of Chemical Engineering Progress}, 48(2):89--94, 1952.

\bibitem{Reichelt}
W.~Reichelt.
\newblock Zur berechnung des druckverlustes einphasig durch- stromter kugel-
  und zylinderschuttungen.
\newblock {\em Rheological Acta}, 17:676--692, 1978.

\bibitem{Felice}
R.~Di Felice and L.G. Gibilaro.
\newblock Wall effects for the pressure drop in fixed beds.
\newblock {\em Chemical Engineering Science}, 59(14):3037--3040, 2004.

\bibitem{Cheng}
N.S. Cheng.
\newblock Wall effect on pressure drop in packed beds.
\newblock {\em Powder Technology}, 210(3):261--266, 2011.

\bibitem{Eisfeld}
B.~Eisfeld and K.~Schnitzlein.
\newblock The influence of confining walls on the pressure drop in packed beds.
\newblock {\em Chemical Engineering Science}, 56:4321--4329, 2001.

\bibitem{KTA}
K.~Ausschusses and KTA.
\newblock Reactor core design of high-temperature gas-cooled reactors - part 3:
  Loss pressure through friction pebble bed cores.
\newblock 1981.

\bibitem{HassanKang}
Y.~Hassan and C.~Kang.
\newblock Pressure drop in a pebble bed reactor under high reynolds number.
\newblock {\em Nuclear Technology}, 180(2), 2012.

\bibitem{Nguyen}
T.~Nguyen, R.~Muyshondt, and Y.~Hassan.
\newblock Experimental investigation of cross flow mixing in a randomly packed
  bed and streamwise vortex characteristics using particle image velocimetry
  and proper orthogonal decomposition analysis.
\newblock {\em Physics of Fluids}, 31(2), 2019.

\bibitem{kenig}
E.Y. Kenig and T.~Atmakidis.
\newblock Cfd-based analysis of the wall effect on the pressure drop in packed
  beds with moderate tube/particle diameter ratios in the laminar flow regime.
\newblock {\em Chemical Engineering Journal}, 155(1-2):404--410, 2009.

\bibitem{Das}
S.~Das, N.G. Deen, and J.A.M Kuipers.
\newblock A dns study of flow and heat transfer through slender fixed-bed
  reactors randomly packed with spherical particles.
\newblock {\em Chemical Engineering Science}, 160:1--19, 2017.

\bibitem{Jun}
S.~Jun, Z.~Yanhua, and L.~Fu.
\newblock Various bypass flow paths and bypass flow ratios in htr-pm.
\newblock {\em Energy Procedia}, 39:258--266, 2013.

\bibitem{Yildiz}
M.~A. Yildiz, G.~Botha, H.~Yuan, E.~Merzari, R.~Kurwitz, and Y.~A. Hassan.
\newblock Direct numerical simulation of the flow through a randomly packed
  pebble bed.
\newblock {\em Journal of Fluids Engineering}, 142(4), 2020.

\bibitem{NOVAK2021107968}
AJ~Novak, S~Schunert, RW~Carlsen, P~Balestra, RN~Slaybaugh, and RC~Martineau.
\newblock Multiscale thermal-hydraulic modeling of the pebble bed
  fluoride-salt-cooled high-temperature reactor.
\newblock {\em Annals of Nuclear Energy}, 154:107968, 2021.

\bibitem{fischer2016}
Paul Fischer, J~Lottes, S~Kerkemeier, O~Marin, K~Heisey, E~Obabko, E~Merzari,
  and Y~Peet.
\newblock Nek5000 user documentation.
\newblock {\em Argonne National Laboratory, Lemont, IL, Report No.
  ANL/MCS-TM-351}, 2016.

\bibitem{merzari2020}
Elia Merzari, Paul Fischer, Misun Min, Stefan Kerkemeier, Aleksandr Obabko,
  Dillon Shaver, Haomin Yuan, Yiqi Yu, Javier Martinez, Landon Brockmeyer,
  et~al.
\newblock Toward exascale: overview of large eddy simulations and direct
  numerical simulations of nuclear reactor flows with the spectral element
  method in nek5000.
\newblock {\em Nuclear Technology}, 206(9):1308--1324, 2020.

\bibitem{Cardinal}
E.~Merzari, H.~Yuan, M.~Min, D.~Shaver, R.~Rahaman, P.~Shriwise, P.~Romano,
  A.~Talamo, Y.~Lan, D.~Gaston, R.~Martineau, P.~Fischer, and Y.~A. Hassan.
\newblock Cardinal: A lower length-scale multiphysics simulator for pebble-bed
  reactors.
\newblock {\em Nuclear Technology}, 2021.

\bibitem{merzari2021}
Elia Merzari, Haomin Yuan, Misun Min, Dillon Shaver, Ronald Rahaman, Patrick
  Shriwise, Paul Romano, Alberto Talamo, Yu-Hsiang Lan, Derek Gaston, et~al.
\newblock Cardinal: A lower length-scale multiphysics simulator for pebble-bed
  reactors.
\newblock {\em Nuclear Technology}, pages 1--23, 2021.

\bibitem{PronghornTheoryManual}
A.~Novak, S.~Schunert, R.~Carlsen, P.~Balestra, D.~Andrs, J.~Kelly,
  R.~Slaybaugh, and R.~Martineau.
\newblock {\em Pronghorn Theory Manual}, 2020.

\end{thebibliography}

\end{document}